\documentclass{elsart}

\usepackage{amsmath,amssymb,graphicx,bm}

\newcommand{\beqn}{\begin{equation}}
\newcommand{\eeqn}{\end{equation}}
\newcommand{\be}{\begin{equation}}
\newcommand{\ee}{\end{equation}}
\newcommand{\bea}{\begin{eqnarray}}
\newcommand{\eea}{\end{eqnarray}}
\newcommand{\ba}{\begin{align}}
\newcommand{\ea}{\end{align}}

\newcommand{\lm}{\Lambda}

\newcommand{\Vlk}{V_{{\rm low}\,k}}

\newcommand{\Tlk}{T_{{\rm low}\,k}}

\newcommand{\fmi}{\, \text{fm}^{-1}}

\begin{document}

\begin{frontmatter}

\title{Variational Calculations \\ using Low-Momentum Potentials
          \\  with Smooth Cutoffs}
\author{S.K.\ Bogner}
\ead{bogner@mps.ohio-state.edu}
and
\author{R.J.\ Furnstahl}
\ead{furnstahl.1@osu.edu}

\address{Department of Physics,
         The Ohio State University, Columbus, OH\ 43210}

\date{\today}
%

\begin{abstract}
%
Recent
variational calculations of the deuteron and the triton illustrate
that simple wave function ans\"atze become more effective 
after evolving the nucleon-nucleon potential to
lower momentum (``$\Vlk$'').
However, wave function artifacts from the use of sharp cutoffs in relative
momentum decrease effectiveness for small cutoffs 
($< 2\,\mbox{fm}^{-1}$) 
and slow down convergence in harmonic oscillator bases.
These sharp cutoff artifacts are eliminated when $\Vlk$ is generated 
using a sufficiently smooth cutoff regulator.

\end{abstract}

\end{frontmatter}

\section{Introduction}

Variational calculations for nuclei  are complicated by the strong short-range
repulsion and tensor forces of conventional nucleon-nucleon potentials,  which
necessitate highly correlated trial wave functions. However, the
nonperturbative nature of conventional inter-nucleon interactions is strongly
scale or resolution dependent and can be radically modified by using the
renormalization group (RG) to lower the momentum cutoff  of the two-nucleon
potential \cite{perturbative}.   A particular consequence is that
the short-range correlations in few- and many-body wave functions are
significantly reduced at lower resolutions \cite{perturbative}.   
This has the practical implication that variational calculations should be more
effective at lower cutoffs using simple wave function ans\"atze. A recent study
using low-momentum potentials (generically called ``$\Vlk$'') supports this
claim \cite{Bogner:2005fn}. This conclusion is also consistent with the results
of Viviani et al.\ \cite{Viviani:2005gu}, 
who showed that variational calculations of 3-- and 4--body
nuclei using hyperspherical harmonics converge faster for potentials with
greater nonlocality, which reduces short-range correlations. 

The optimistic conclusions regarding the use of $\Vlk$ in variational
calculations are clouded by some problems associated with the sharp momentum
cutoff, which are of concern for some practical applications. In particular, one
would expect that for very-low-energy observables  such as the deuteron or
triton binding energies, there should be improved  variational estimates down
to very low cutoffs (e.g., comparable to the ``binding momentum''), and rapid
convergence  with basis size. However, what was observed in
Ref.~\cite{Bogner:2005fn} was improvement down to moderate cutoffs of
$2\,\mbox{fm}^{-1}$ followed by a degradation of  the variational estimates at
smaller cutoffs.  Further, for a wide range of cutoffs the  convergence of the
triton binding energy with the size of a harmonic oscillator basis was
exceedingly slow once the energy prediction was at the 100\,keV level. Both of
these problems were attributed to  the use of a sharp cutoff on the relative
momentum.  In this letter, we  revisit the variational calculations in
Ref.~\cite{Bogner:2005fn} to  demonstrate that these limitations 
can be eliminated
by constructing $\Vlk$ potentials using a smooth regulator. There are many new
issues to consider with such a regulator,  but we defer most of the further
discussion to a more complete
and wide-ranging investigation \cite{Bogner2006b}.


\section{$\Vlk$ with a Smooth Regulator}
 \label{sect:smooth}

The construction of $\Vlk$ with a sharp cutoff is documented in
Refs.~\cite{VlowkRG,VLOWK}, where it is shown that either
RG equations or Lee-Suzuki transformations
can be used.  
The latter approach relies heavily on the introduction of
orthogonal low- and high-energy subspaces with projection operators
$P$ and $Q$, such that $P + Q = 1$ and $P Q = Q P = 0$.
In momentum space for the two-nucleon system, the last condition mandates 
a sharp cutoff $\Lambda$ in relative momentum, so that $P$-space integrals run
from 0 to $\Lambda$ while $Q$-space integrals run from $\Lambda$ to $\infty$
(or to a large bare cutoff). 
But while 
replacing a sharp cutoff with a regulator that smoothly cuts off
the relative momentum seems incompatible with methods requiring $PQ = 0$,
it is not a conceptual problem for the more general RG approach~\cite{VlowkRG}. 
Details will be presented in Ref.~\cite{Bogner2006b}; here we simply summarize
a three-step procedure applied in the present calculations.

Smooth cutoff
regulators will be applied in each partial wave as simple functions
of the relative momentum.
It is convenient and efficient for numerical calculations to
define the smoothly regulated energy-independent potential $\Vlk$ and 
the corresponding $\Tlk$ matrix in terms of a reduced potential $v$
and a reduced $T$ matrix $t$ as
\be
  \Vlk(k',k) = f(k')v(k',k)f(k) \ ,
\ee
and
\be
  \Tlk(k',k;k^2) = f(k')t(k',k;k^2)f(k) \ ,
  \label{define}
\ee
where $f(k)$ is a smooth cutoff function.  Here we adopt for
$f(k)$ the exponential form used in 
chiral EFT potentials at N$^3$LO order \cite{N3LO},
\beqn
   f(k) = e^{-(k^2/\lm^2)^n} \ .
   \label{eq:expreg}
\eeqn
We use $n=4$ throughout this work.
The reduced $t$ matrix obeys a Lipmann-Schwinger 
equation with loop integrals smoothly cut off by the internal factors of
$f(p)$ \cite{VlowkRG},
\beqn
  t(k',k;k^2) = v(k',k) 
     + \frac{2}{\pi}\int_{0}^{\infty}\!p^2dp
            \frac{v(k',p)f^2(p)t(p,k;k^2)}{k^2-p^2} \ .
\eeqn
Note that the cutoff is on the loop momentum 
but not on external momenta. Principal
value integrals are implicit throughout.  

In the energy-independent RG approach, 
an RG equation for the reduced
interaction $v(k',k)$ 
is derived by demanding 
that $\frac{d}{d\Lambda}t(k',k;k^2)=0$.%
\footnote{This RG equation generates a non-hermitian interaction.
Hermiticity can be restored by using a symmetrized form of the RG
equation~\cite{Bogner2006b}. The resulting  interaction preserves the on-shell
$T$ matrix, but no longer preserves the  half-offshell $T$ matrix.} 
Using the large-cutoff  initial condition $v(k',k)=V_{NN}(k',k)$, the resulting
set of coupled differential equations can be numerically integrated to evolve
the interaction to lower cutoffs. The  resulting $\Vlk$ preserves the original
on-shell $T$ matrix up to factors of the smooth regulator,
\be
\label{tmatequiv}
\Tlk(k,k;k^2)=f^2(k)T_{NN}(k,k;k^2), 
\ee
which implies that low-energy phase shifts are preserved for on-shell
momenta away from the transition region near the cutoff where the
regulator function rapidly decreases to zero. While the energy-independent RG approach 
provides a direct path to construct the smooth cutoff version of $\Vlk$, it is not the most convenient
or numerically robust method \cite{Bogner2006b}. 

In this work, we prefer to use a much simpler \emph{energy-dependent} RG equation
to evolve the bare potential $V_{\rm NN}$ to a lower
cutoff. The energy-dependent RG equation is obtained 
by requiring invariance
of the \emph{full} off-shell $T$ matrix, 
$\frac{d}{d\Lambda}t(k',k;E)=0$, 
which can be formally integrated~\cite{Bogner2006b} to
recover the Bloch-Horowitz equation with a smooth cutoff:
\beqn
    v(k',k;E) = V_{\rm NN}(k',k)
       + \frac{2}{\pi}\int_{0}^{\infty}\!p^2dp 
           \biggl(1-f^2(p)\biggr)
	   \frac{V_{\rm NN}(k',p)v(p,k;E)}{E-p^2}
	\ .
\eeqn 
The resulting integral equation for $v(k',k;E)$ is much more efficient
for numerical calculations than the original set of
coupled differential equations.

The second step is to trade the energy dependence for momentum dependence by
defining an energy-independent (but non-hermitian) $\Vlk(k',k)$ that gives the same half-on-shell $T$ matrix
and wave functions as the energy-dependent interaction $V_{\rm eff}(k',k;E)=f(k')v(k',k;E)f(k)$,
\beqn
   \langle k'\!\mid\! V_{\rm eff}(p^2)\!\mid\!\Psi_p\rangle
     =\langle k'\!\mid\! \Vlk\!\mid\!\Psi_p\rangle
     \ .
\eeqn
The $\Psi_p$ are the self-consistent wave functions of the energy-dependent low-momentum
Hamilitonian $H_{\rm eff}(p^2)$. Using the completeness of these wave functions, one 
obtains a simple expression for the non-hermitian $\Vlk$,
\beqn
  \Vlk(k',k) = \biggl( \frac{2}{\pi} \biggr)^2
    \int_{0}^{\infty}\! p^2dp
    \int_{0}^{\infty}\! k''^2 dk''\, V_{\rm eff}(k',k'';p^2)
            \Psi_p(k'')\widetilde{\Psi}_p\!^*(k)
        \ ,
\eeqn
where $\tilde{\Psi}_p^*$ is the biorthogonal complement wave function. Note
that the integral over the continuous scattering states will include a
summation  over discrete bound states, when present. 

The final step is to apply a Gram-Schmidt procedure to 
hermitize the potential,
as prescribed in Ref.~\cite{Kuoherm}. The end result is a hermitian,
energy-independent $\Vlk$ with a smooth cutoff regulator that preserves the low
energy on-shell $T$ matrix up to factors  of the regulator as in
Eq.~(\ref{tmatequiv}). 
In addition, the complete set of wave
functions obtained from diagonalizing $H_{\text{low k}}$ can be used to
consistently evolve general operators beyond the Hamiltonian with the smooth
cutoff.   Thus, the present approach may be viewed as a generalization of
conventional effective interaction methods such as Lee-Suzuki transformations
to smooth cutoffs.  Further details concerning the present  approach as well as
the energy-independent RG method will be provided in Ref.~\cite{Bogner2006b}.

As in Ref.~\cite{Bogner:2005fn},
we will show results starting from the Argonne
$v_{18}$ potential~\cite{AV18}
as $V_{\rm NN}$, since it has been used in almost all modern calculations
of light nuclei.  
However, as with the sharp cutoff $\Vlk$ calculations, the pattern of
results for the full cutoff range shown here does not vary
significantly with different initial potentials.
When using the sharp cutoff $\Vlk$,
two-body bound-state properties and phaseshifts 
are preserved by construction for external relative 
momenta right up to the cutoff.
(Note that 
three- and many-body observables require the consistent addition of a
three-body force to remove cutoff dependence \cite{VLOWKFEW}.)
The use of a smooth regulator, however, distorts the phaseshifts
near the cutoff according to Eq.~(\ref{tmatequiv})
to a degree that depends on the type of regulator function \cite{Bogner2006b}.
These distortions are not important for the low-energy observables discussed
here, but will need further assessment for future applications.

Since the final $\Vlk$ is energy independent and hermitian, 
variational calculations proceed as described in ordinary quantum
mechanics texts (e.g., without special normalizations as needed for
energy-dependent potentials).  
That is, given a trial wave function $\psi_{\rm trial}$, our variational
estimate for the ground state energy at cutoff $\Lambda$ is:
\be
   E_{\rm var}(\Lambda) =
    \frac{\langle \psi_{\rm trial}| T + \Vlk(\Lambda) | \psi_{\rm trial}\rangle}
         {\langle \psi_{\rm trial}|\psi_{\rm trial}\rangle}
         \ ,
\ee
which we minimize with respect to the parameters in $\psi_{\rm trial}$.
Alternatively, we get a variational estimate by diagonalizing 
$T + \Vlk(\Lambda)$ in a truncated basis, where the trial wavefunction
is a linear combination of the basis functions.


\section{Variational Results for the Deuteron and Triton}
  \label{sect:results}

Here we retrace the calculations of Ref.~\cite{Bogner:2005fn},
starting with a study of the deuteron binding energy.
As noted there, for weakly bound states we expect
that a simple, generic ansatz should work 
increasingly well as the cutoff is lowered.
Two such ans\"atze were considered.
In the first one \cite{SALPETER51},   
the (unnormalized) $^3S_1$ and $^3D_1$ trial functions for the deuteron
are (following the conventions of Ref.~\cite{MACHLEIDT01})
\be
  \psi_0(k) = \frac{1}{(k^2 + \gamma^2)(k^2 + \mu^2)}  \ , 
  \qquad
  \psi_2(k) =  \frac{a\, k^2}{(k^2 + \gamma^2)(k^2 + \nu^2)^2} \ ,
  \label{eq:salpeter}
\ee 
where $\gamma$, $\mu$, $\nu$, and $a$ are variational parameters.  
The underlying physics implies that $\mu$ and $\nu$ should 
be roughly the inverse range of the interaction
and $\gamma$ should be close to $(-M_N E_d)^{1/2}$, where
$M_N$ is the mean neutron-proton mass and $E_d \approx
-2.2246\,\mbox{MeV}$ is the deuteron binding energy. 
The regulator 
in $\Vlk$ implies that the corresponding
exact deuteron wave function does not
contain high-momentum components. Therefore, the two-body trial wave functions
are multiplied by the same regulator $f(k)$ for the relative momentum.

\begin{figure}[t]
\centerline{\includegraphics*[width=3.7in]{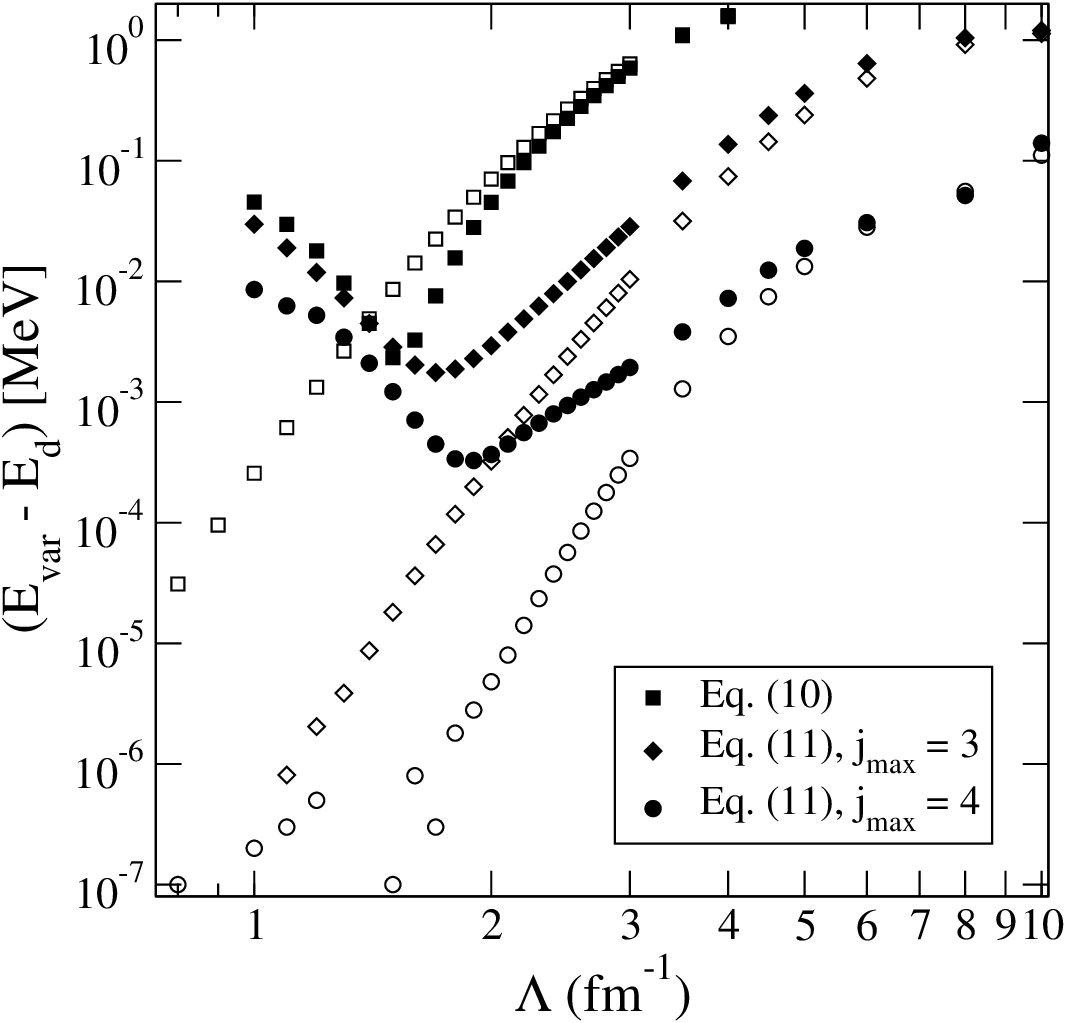}}
\vspace*{-.1in}
\caption{Deviation from $E_d$ of the best variational energy
as a function of cutoff $\Lambda$ for the wave function ans\"atze of
Eqs.~(\ref{eq:salpeter}) and (\ref{eq:machleidt}) 
with different numbers of terms. Smooth cutoff $\Vlk$ results are given by the open
symbols.}
\label{fig:deuteron}

\end{figure}

We also adapted the form used for a high-accuracy representation 
of the deuteron
wave function in Ref.~\cite{MACHLEIDT01}, for which
\be
  \psi_0(k) = f(k) \sum_{j=1}^{j_{\rm max}} \frac{C_j}{k^2 + m_j^2} \ ,
  \qquad
  \psi_2(k) = f(k) \sum_{j=1}^{j_{\rm max}} \frac{D_j}{k^2 + m_j^2} \ ,
  \label{eq:machleidt}
\ee
where the $m_j$ are fixed in a geometric progression:
\be 
  m_j = (-M_N E_d)^{1/2} + (j-1)m_0\ , \quad \mbox{with}\ 
  m_0 = 0.9\,\mbox{fm}^{-1} \ ,
\ee
by treating the $C_j$ and $D_j$ coefficients
as variational parameters for a given value of
$j_{\rm max}$.  
(The very accurate \emph{parameterization} of the deuteron wave function for
the Bonn potential in Ref.~\cite{MACHLEIDT01} has $j_{\rm max} = 11$, with some
constraints on the $C_j$'s and $D_j$'s.)
Since the variational coefficients appear linearly, we can simply
diagonalize the Hamiltonian in the truncated basis of Eq.~(\ref{eq:machleidt})
to find the best estimate of the deuteron energy.

The best variational energy for Eq.~(\ref{eq:salpeter}) 
as a function of a sharp cutoff is shown as the filled squares
in Fig.~\ref{fig:deuteron}.
These estimates are not even bound for cutoffs above
$\Lambda \approx 5\, \mbox{fm}^{-1}$ (which includes the bare Argonne $v_{18}$
potential) but
rapidly improve as the cutoff is lowered further, reaching
a minimum deviation of less than 3\,keV around $\Lambda
\approx 1.5\,\mbox{fm}^{-1}$. Similar results are found for the ansatz of Eq.~(\ref{eq:machleidt}) 
with $j_{\rm max}=3$ (solid diamonds) and $j_{\rm max}=4$ (solid circles). 
As expected, lowering the sharp cutoff dramatically improves the effectiveness of the 
simple wave function ans\"atze, 
but the results unexpectedly worsen for cutoffs that 
are significantly larger than the naive limiting value set
by the ``binding momentum'' of the deuteron, $\Lambda_d\approx 0.25 \fmi$. 

\begin{figure}[t]
\centerline{\includegraphics*[width=4.1in,angle=0]{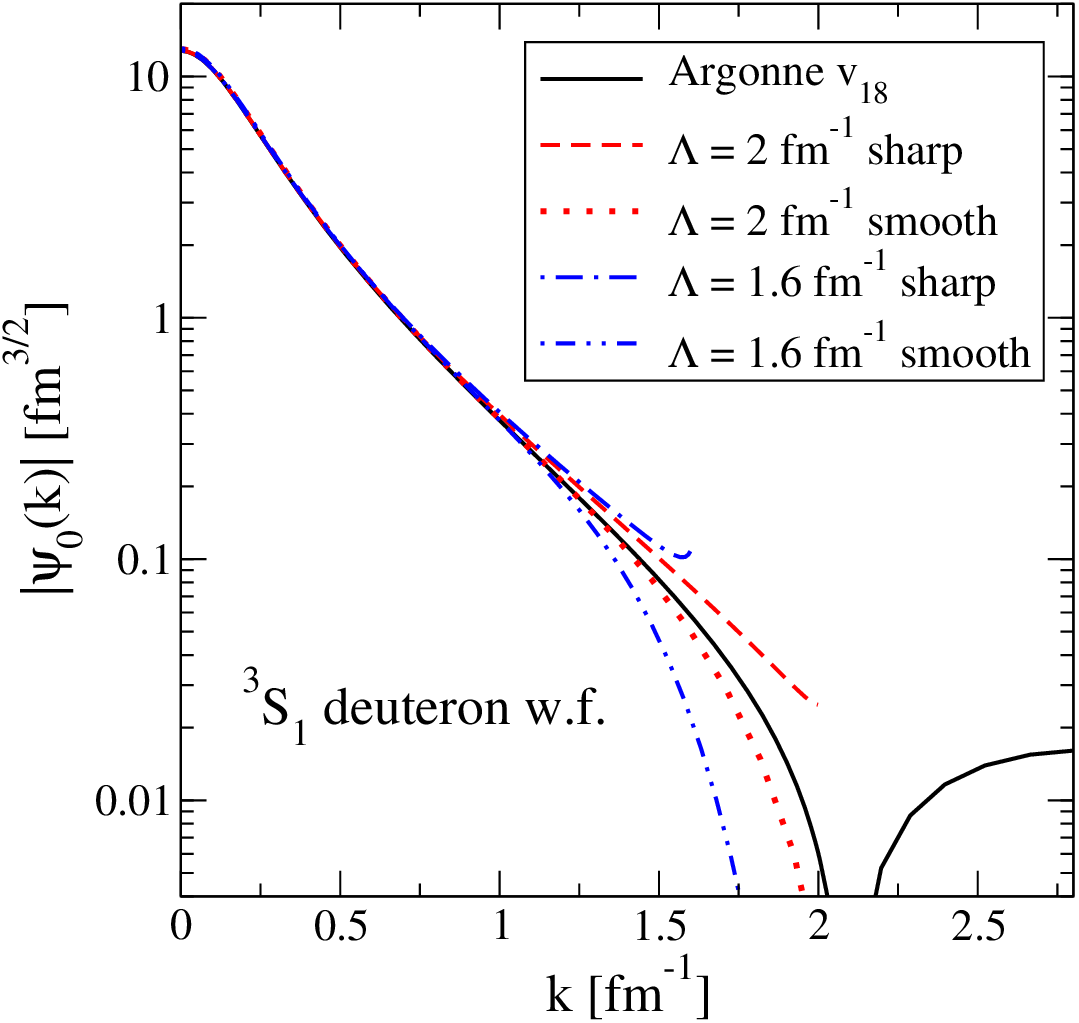}}
\vspace*{-.1in}
\caption{Momentum-space $^3S_1$ deuteron wave function for two 
    cutoffs and for the bare Argonne $v_{18}$ potential.}
\label{fig:kspacewf}
\end{figure}        

The variational estimates 
for the smooth cutoff of Eq.~(\ref{eq:expreg}) 
are shown in Fig.~\ref{fig:deuteron}
as unfilled symbols for the corresponding
ans\"atze.
For the ansatz of Eq.~(\ref{eq:salpeter}), the smooth cutoff results are inferior
at intermediate cutoffs but continue to improve monotonically at smaller cutoffs.
For the other ansatz, the smooth-cutoff
results are superior throughout and 
improve until reaching agreement with the exact result at the eV level.
This is in accord with intuition for such a low-energy bound state,
and it emphasizes that one works much too hard in calculating low-energy
observables using conventional potentials that contain
strong high-momentum components.

The reduced effectiveness 
of simple ans\"atze for the deuteron when using sharp cutoffs 
below $2\,\mbox{fm}^{-1}$ can
be understood by
looking at the corresponding wave functions in momentum space.      
In Fig.~\ref{fig:kspacewf}, we show the exact deuteron
wave functions in momentum space for both the smooth and sharp
cutoffs at $\Lambda = 1.6\,\mbox{fm}^{-1}$ and $2\,\mbox{fm}^{-1}$.
The $\Vlk$ wave functions remove the
short-range/high-momentum behavior (e.g., the node just
above $2\,\mbox{fm}^{-1}$ \cite{Garcon:2001sz}) that is increasingly resolved at
higher cutoffs, requiring finer and finer cancellations in the 
variational integrals and a more correlated wave function.
At $\Lambda = 2\,\mbox{fm}^{-1}$, the sharp wave function is well behaved
and an ansatz cutoff at the same momentum is adequate for a momentum-space
variational calculation.  
(Even here, the abrupt cutoff creates problems in coordinate-space 
calculations, particular for the $^3D_1$ component of the wave function.)
But by 
$\Lambda = 1.6\,\mbox{fm}^{-1}$, one clearly sees a complicated endpoint behavior
that will not be reproduced in simple variational trial functions.
In contrast, the smooth regulator potential and
corresponding wave functions do not have these artifacts.

The extension from the deuteron to the triton in Ref.~\cite{Bogner:2005fn}
was kept simple by using a truncated harmonic oscillator
basis for a variational 
calculation with the two-body interaction only, 
which we repeat here for the smoothly regulated $\Vlk$ potentials.
The antisymmetric three-nucleon basis is generated from
the Jacobi coordinate oscillator states~\cite{Navratil}
\be
\mid\!(nlsjt;\mathcal{NL}\frac{1}{2}\mathcal{J}\frac{1}{2})JT\rangle,
\label{eq:jacobi}
\ee 
where $(nlsjt)$ and $(\mathcal{NL}\frac{1}{2}\mathcal{J}\frac{1}{2})$ are the quantum numbers
corresponding to the two relative Jacobi coordinates [e.g., 
${\bf k}=\frac{1}{2}({\bf p}_1-{\bf p}_2)$ and
${\bf q}=\frac{2}{3}({\bf p}_3-\frac{1}{2}({\bf p}_1+{\bf p}_2))$], 
and the basis is
truncated according to the total number of oscillator quanta $N=(2n+l+2\mathcal{N}+\mathcal{L})\le N_{max}$. 
Diagonalizing the intrinsic Hamiltonian in the truncated basis
and minimizing with respect to the oscillator 
length parameter $b$ provides a variational estimate to the true ground-state 
energy.

\begin{figure}[t]
\centerline{\includegraphics*[width=4.50in]{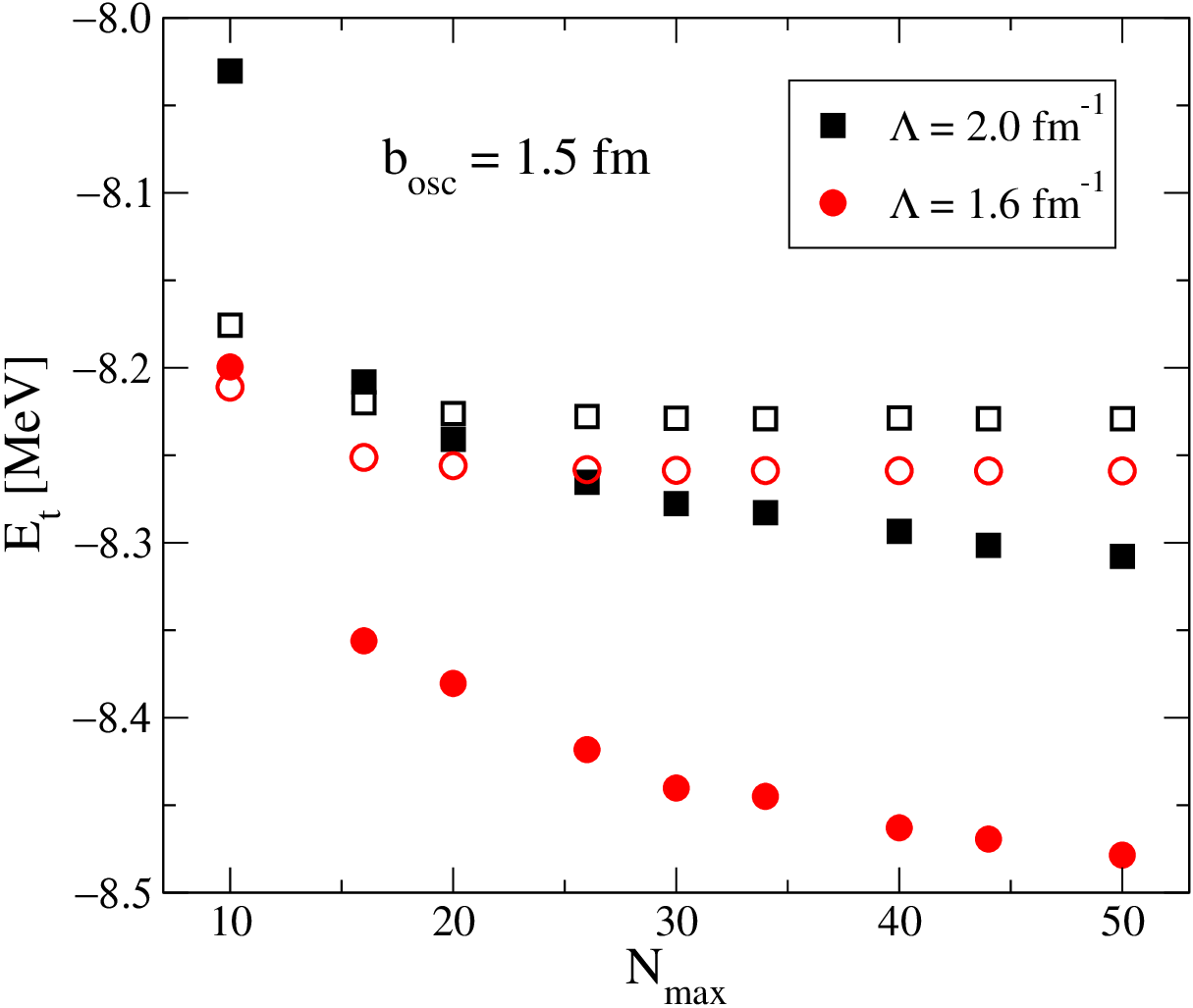}}
\vspace*{-.1in}
 \caption{Convergence of the binding energy of the triton from a direct
    diagonalization of the Hamiltonian in a harmonic-oscillator basis
    with fixed oscillator parameter $b_{\rm osc} = 1.5\,\mbox{fm}$.
    Results for two cutoffs are shown. 
    In each case 
    the filled symbols are for a sharp cutoff and the unfilled symbols
    are for a smooth cutoff.}
\label{fig:triton}
\end{figure}

\begin{figure}[t]
\centerline{\includegraphics*[width=4.01
in]{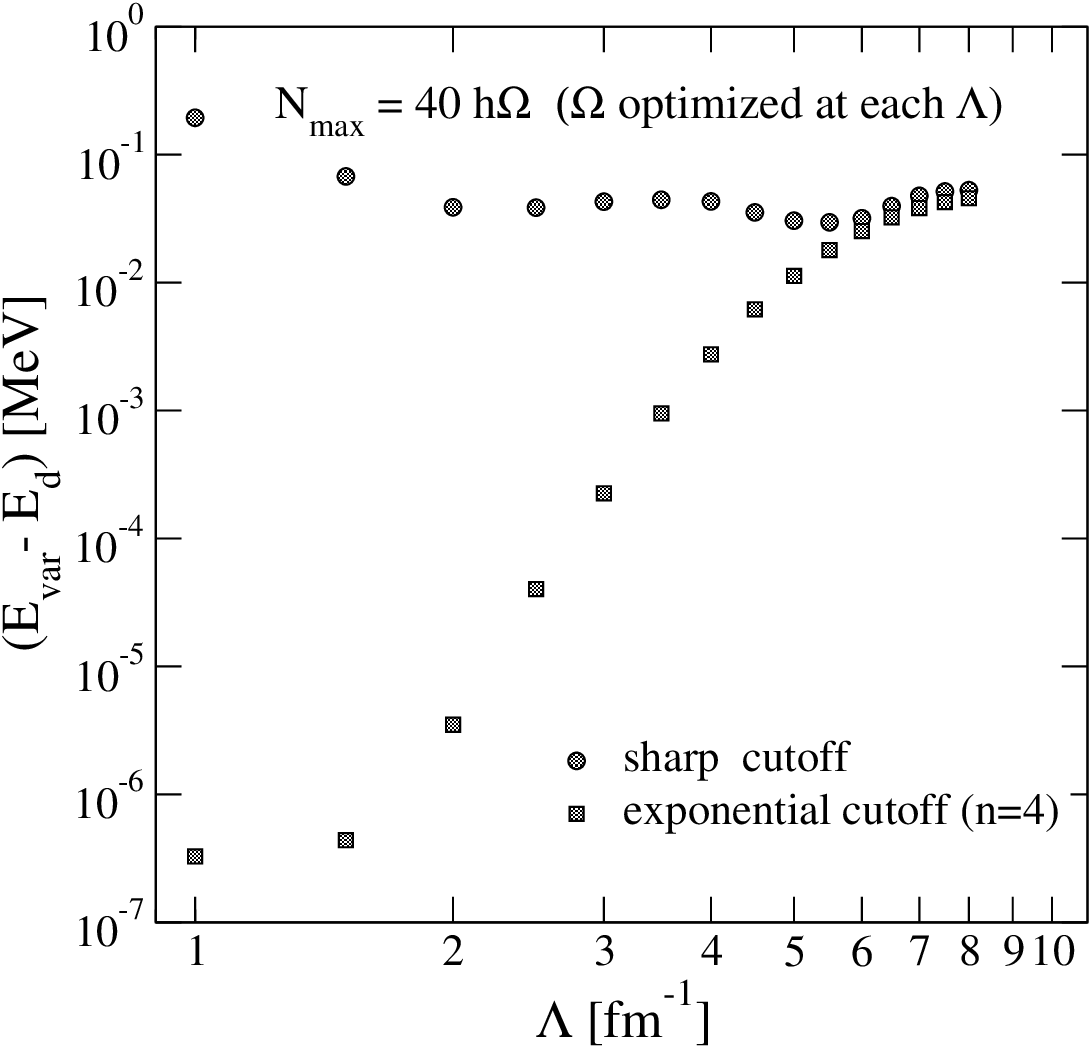}}
\vspace*{-.1in}
\caption{Deviation from $E_d$ of the optimized variational energy
as a function of cutoff $\Lambda$ for a harmonic oscillator basis
of fixed size $N_{\rm max} = 40$, comparing a sharp cutoff to an exponential
regulator
with $n=4$.  The
oscillator parameter is optimized for each $\Lambda$. }
\label{fig:deuteronHO}
\end{figure} 
       
The same pattern of the variational estimates  for sharp cutoffs seen for the
deuteron,  namely improvement to a minimum followed by worsening estimates for
very low $\Lambda$, was observed for the triton in Ref.~\cite{Bogner:2005fn}.
However,  an additional problem of convergence with the size of the  harmonic
oscillator basis, which is of greater practical importance, was seen as well
over a wide range of cutoffs. For cutoffs around $2\,\mbox{fm}^{-1}$, energies
within about 200\,keV of the accurate result from Fadeev calculations are
reached with relatively small basis size, but then further convergence as the
basis is increased is extremely slow (see the filled symbols in
Fig.~\ref{fig:triton}). In fact, extremely slow convergence beyond the $50$\,
keV level is found for a wide range of cutoffs. One expects convergence at the
keV level for good variational calculations \cite{Viviani:2005gu}.

The poor convergence of the sharp cutoff $\Vlk$ in an oscillator basis 
is analogous to ``Gibbs overshoot'' phenomena found in
Fourier series expansions of discontinuous functions. 
This connection is supported by noting the rapid convergence
of the energy with the smooth cutoff 
$\Vlk$ (e.g., the unfilled squares for $\Lambda=2.0 \fmi$
in Fig.~\ref{fig:triton} are
within 2 keV at $N_{\rm max} = 20$ and within 1 keV at
$N_{\rm max} = 30$  of the converged result). 
Similar dramatically improved convergence using the smooth cutoff  
$\Vlk$ is also found at other cutoffs 
(e.g., the unfilled circles in Fig.~\ref{fig:triton}).

Fadeev results
with smooth cutoffs are not yet available to verify that they agree with
the converged energies obtained here.  
However, we can look back at 
deuteron calculations using a harmonic oscillator 
basis to anticipate what we will find.
Figure~\ref{fig:deuteronHO} shows the deviation of the variational estimate 
from the exact deuteron binding 
energy for a harmonic oscillator basis, comparing
the two types of regulator.  Each point was
optimized with respect to the oscillator parameter in
a basis of fixed size $N_{\rm max} = 40$.  As found for the other ans\"atze,
the smooth cutoff improves steadily with decreasing $\Lambda$.
We expect similar behavior for the triton.%
\footnote{We do not expect the addition of consistent three-body forces,
which have been shown to become more perturbative with lower 
cutoffs~\cite{VLOWKFEW,perturbative}, to
alter this assessment.}

\section{Conclusions}  

In summary,
use of a smooth cutoff regulator preserves the improvement with
lower cutoff and remedies the technical problems noted
in Ref.~\cite{Bogner:2005fn}
for variational calculations of the deuteron and triton
with low-momentum potentials.
While the deuteron and triton are not stringent tests for heavier nuclei,
these results 
coupled with the
rapid convergence of the particle-particle channel 
observed in nuclear matter \cite{perturbative}
imply that low-momentum
potentials 
will be much more effective for few-body and many-body 
variational calculations than any conventional large-cutoff potential.
Investigations of the effectiveness of running a smooth cutoff lower for
chiral EFT potentials, which are themselves
low-momentum potentials compared to conventional potential models, 
will be explored elsewhere 
along with many other issues \cite{Bogner2006b}. 

To take advantage of these observations,
the variational calculations described recently in Ref.~\cite{Viviani:2005gu},
which show the advantages of nonlocal interactions, should
be particularly well suited.
Based on our results here,
we anticipate even more efficient variational
results for low-momentum interactions with smooth cutoffs, 
with the added advantage of
being able to vary the cutoff as a tool to optimize and probe the quality of the 
solution.
Furthermore,
we can avoid the problem of constructing consistent, model-independent operators
for conventional potentials by concurrently evolving to low momentum the potential and operators
from chiral EFT.

Since Hartree-Fock becomes a reasonable starting point for nuclear many-body
calculations \cite{perturbative},  the large arsenal of techniques developed
for the Coulomb many-body problem becomes available and should be explored as
well. In addition, the development of the smooth cutoff $\Vlk$  allows one to
unambiguously perform a density matrix expansion starting from low-momentum 
interactions to gain insight into the microscopic foundation of the nuclear
energy density functional, as the required coordinate space quantities are now
well-defined with the smooth cutoff \cite{DME}.

\begin{ack}
We thank Sunethra Ramanan and Achim Schwenk
for useful comments and discussions.
This work was supported in part by the National Science Foundation
under Grant No.~PHY--0354916.
\end{ack}


\end{document}